\documentclass[conference]{IEEEtran}
\IEEEoverridecommandlockouts
\usepackage{cite}
\usepackage{amsmath,amssymb,amsfonts}
\usepackage{algorithmic}
\usepackage{pstricks}
\usepackage{graphicx}
\usepackage{textcomp}
\usepackage{xcolor}
\usepackage{multirow}
\usepackage{bbm}
\usepackage{dsfont}
\usepackage{comment}
\usepackage{breqn}
\usepackage[linesnumbered,ruled,vlined]{algorithm2e}
\usepackage{caption}
\usepackage{subcaption}
\usepackage{multicol}
\usepackage{eso-pic}

\usepackage{adjustbox}
\newcommand\AtPageUpperMyright[1]{\AtPageUpperLeft{%
 \put(\LenToUnit{0.5\paperwidth},\LenToUnit{-1cm}){%
     \parbox{0.5\textwidth}{\raggedleft\fontsize{9}{11}\selectfont #1}}%
 }}%
\newcommand{\conf}[1]{%
\AddToShipoutPictureBG*{%
\AtPageUpperMyright{#1}
}
}
\usepackage{mathrsfs}
\usepackage[scr=rsfs,cal=boondox]{mathalfa}
\newcommand{\nonl}{\renewcommand{\nl}{\let\nl\oldnl}}
\def\BibTeX{{\rm B\kern-.05em{\sc i\kern-.025em b}\kern-.08em
    T\kern-.1667em\lower.7ex\hbox{E}\kern-.125emX}}

\begin{document}

\title{Performance Modeling Sparse MTTKRP \\ Using Optical Static Random Access Memory \\ on FPGA}

\author{
    \IEEEauthorblockN{Sasindu Wijeratne\IEEEauthorrefmark{1}, Akhilesh Jaiswal\IEEEauthorrefmark{2}, Ajey P. Jacob\IEEEauthorrefmark{2}, Bingyi Zhang\IEEEauthorrefmark{1},  Viktor Prasanna\IEEEauthorrefmark{1}}
    \IEEEauthorblockA{\IEEEauthorrefmark{1}Department of Electrical and Computer Engineering, University of Southern California, Los Angeles, CA, USA}
    \IEEEauthorblockA{\IEEEauthorrefmark{2}Information Sciences Institute (ISI), University of Southern California (USC), Marina Del Rey,
CA, USA}
    Email: kangaram@usc.edu, \{akjaiswal, ajey\}@isi.edu, \{bingyizh, prasanna\}@usc.edu
}

\conf{26th IEEE High Performance Extreme Computing Conference, 2022}

\maketitle

\begin{abstract}
Electrical static random memory (E-SRAM) is the current standard for internal static memory in Field Programmable Gate Array (FPGA). Despite the dramatic improvement in E-SRAM technology over the past decade, the goal of ultra-fast, energy-efficient static random memory has yet to be achieved with E-SRAM technology. However, preliminary research into optical static random access memory (O-SRAM) has shown promising results in creating energy-efficient ultra-fast static memories. 

This paper investigates the advantage of O-SRAM over E-SRAM in access speed and energy performance while executing sparse Matricized Tensor Times Khatri-Rao Product (spMTTKRP). spMTTKRP is an essential component of tensor decomposition algorithms which is heavily used in data science applications. The evaluation results show O-SRAMs can achieve speeds of 1.1$\times$ - 2.9$\times$ while saving 2.8$\times$ - 8.1$\times$ energy compared to conventional E-SRAM technology. 
\end{abstract}


\section{Introduction}

Recent advances in collecting and analyzing large datasets have led to the information being naturally represented as higher-order tensors. Tensor Decomposition transforms input tensors to a reduced latent space which can then be leveraged to learn salient features of the underlying data distribution. Tensor Decomposition has been successfully employed in many fields, including machine learning, signal processing, and network analysis~\cite{mondelli2019connection,cheng2020novel,wen2020tensor}. Canonical Polyadic Decomposition (CPD) \cite{12041586} is the most popular method of decomposing a tensor to a low-rank tensor decomposition model. It has become the standard tool for unsupervised multiway data analysis. The Matricized Tensor Times Khatri-Rao product (MTTKRP) kernel\cite{8821030} is known to be the computationally intensive kernel in CPD.
Due to the sparse nature of real-world tensors, specialized hardware accelerators are becoming increasingly popular for improving the efficiency of sparse tensor computations. However, memory access time has become the bottleneck in sparse MTTKRP (spMTTKRP) operation due to irregular data access patterns.

The 6 transistor E-SRAM is currently the de-facto standard for on-chip memory storage. However, the memory access speed for E-SRAM is constrained by long electrical wires, constituting the bit-lines and wordlines in an SRAM array and associated parasitic resistances and capacitances. Optical memory systems have the capability to achieve orders of magnitude faster memory access speed using ultra-fast optical signals that do not suffer from fundamental signal transfer speed bottlenecks like their electrical counterparts. Various implementations of optical-SRAM (O-SRAM) have been explored in the past \cite{pleros2008optical, tsakyridis201910, dong2015nano, li2009optical, alexoudi2016iii, pitris2016wdm, liu2006packaged, trita2013monolithic }. However, an O-SRAM technology amenable to existing foundry manufacturing exhibiting ultra-high speed and low-energy consumption has remained challenging. Recently, however, an O-SRAM technology built using foundry-friendly optical devices, with excellent speed and energy-efficiency has been reported in \cite{arxiv.2111.13682}. The O-SRAM reported in \cite{arxiv.2111.13682} consists of an optical bistable element formed by a feedback connection between photodiodes and microring resonators. The bistable optical element can store two levels in a differential manner (i.e. storing the bit and the complement of the bit). Due to its differential nature, similar to E-SRAM, the O-SRAM of \cite{arxiv.2111.13682} features differential read and write ensuring robust memory operations. Further, the photodiodes and the microring resonators are reverse biased ensuring small static current dissipation and hence energy-efficiency compared to previous works. Thus, the recent advances in O-SRAM solutions as in \cite{arxiv.2111.13682}, has driven optical memory system closer to large-scale manufacturing while providing ultra-fast speed and excellent energy-efficiency. It is thereby important to quantify system-level benefits of such emerging optical memory solutions on representative data intensive computational kernels as in Tensor Decomposition.

\begin{figure*}
\centering
\includegraphics[width=\linewidth]{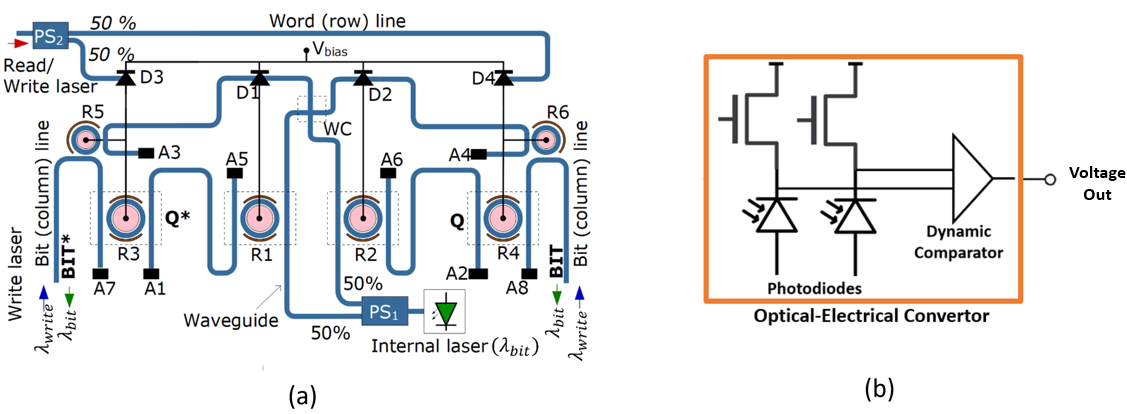}
\caption{(Left) A 2x2 Optical-SRAM array \cite{arxiv.2111.13682} showing cascaded optical bit-cells (Right) electro-optic sense amplifiers used to sense the optical data from memory array and convert it to electrical data}
\vspace{-3mm}
\label{o-sram}
\end{figure*}

Since real-world tensors are sparse, specialized hardware accelerators are attractive for improving the compute efficiency of sparse tensor computations \cite{hicoo_paper, alto_paper, 7161496, 10.1145/3330345.3330366, 10.1145/3295500.3356216}. As spMTTKRP is memory bound, improving the accelerator's internal sustained memory bandwidth and latency can significantly reduce the computation time. Our work mainly focuses on FPGA as it facilitates near memory computing with custom adaptive hardware due to its reconfigurability and large on-chip memory \cite{xilinxalveo, 9286199}. It enables the development of a custom memory hierarchy and compute units.

In this work, we develop a performance model to analyze the acceleration and power efficiency we can achieve by replacing the internal E-SRAMs inside an FPGA with O-SRAMs. The contributions of our paper are as follows:
\begin{itemize}
\item We develop a performance model to analyze the O-SRAM memory design for a FPGA environment.
\item We evaluate the impact of O-SRAM on a memory-bound algorithm by evaluating spMTTKRP using our proposed performance model.
\item Our results show that O-SRAM can improve the total execution time of spMTTKRP by up to 2.9$\times$ compared with traditional E-SRAMs.
\item Our system also shows  2.8$\times$ - 8.1$\times$ energy savings while performing spMTTKRP using O-SRAM compared with traditional E-SRAMs.
\end{itemize}

\begin{figure*}
\centering
\includegraphics[width=0.75\linewidth]{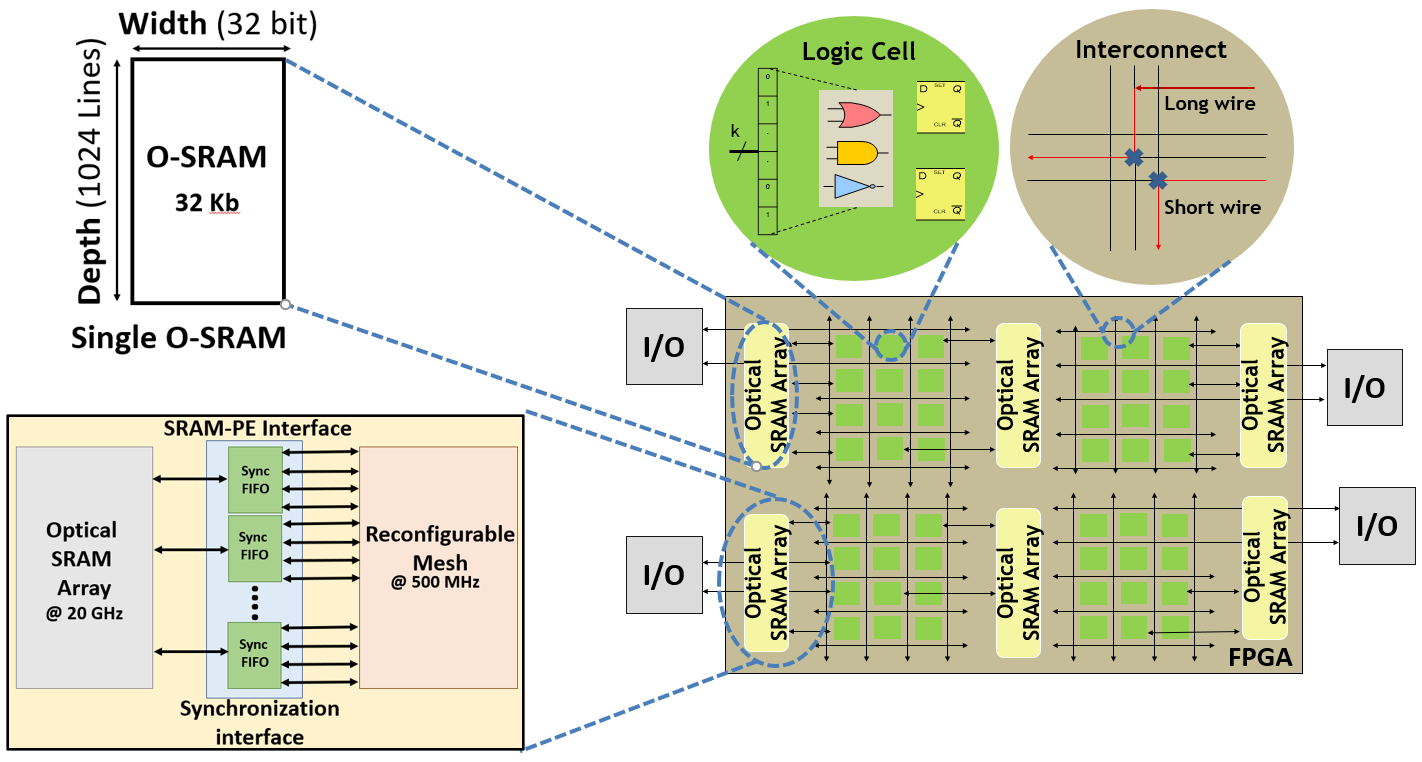}
\caption{Overall FPGA with Optical SRAMs integrated into the architecture}
\label{sram_pipes}
\end{figure*}

\section{Optical Memory Technology} \label{O-technology}

The O-SRAM shown in Fig. \ref{o-sram} is adopted from \cite{arxiv.2111.13682}. It consists of a bistable element formed by photodiodes D1 and D2 and microring resonators (MRRs) R1 and R2. The bistable element is coupled to photodiodes D3-D4 and MRRs R3-R6, which act as read/write access devices. To access a particular data the Wordline waveguide is activated by sending a light pulse through it and the data can be accessed through the waveguides BIT and BIT* in Fig. \ref{o-sram}. Thus, the O-SRAM of Fig. \ref{o-sram} is functionally similar to E-SRAM in the following ways: 1) the data is stored in the bistable element as complementary optical signals, similar to E-SRAM that stores complementary electrical data in a bit-cell 2) the Wordline and BIT/BIT* waveguides are orthogonal to each other, allowing the O-SRAM bit-cell to be cascaded to create a large memory array. 

Due to the optical nature of storage, the operating speed of O-SRAM can be orders of magnitude faster than its electrical counterparts. Further, as the array size increases, the speed of E-SRAM drastically reduces due to parasitic resistance and capacitances associated with metal wires. O-SRAM on the other hand can have very large area arrays without any significant degradation in speed as data is accessed through waveguides that carry optical signals as opposed to metal lines carrying electrical signals. O-SRAM are thus suitable for large-scale chips as in wafer-scale systems since they can transfer ultra-fast data across long distances using optical signaling. Wafer-scale systems are chips built on an entire 300mm wafer \cite{arxiv.2010.03660}. Wafer-scale system also helps to accommodate the high area associated with optical memories. An O-SRAM bit-cell is over three orders of magnitude larger in size compared to an E-SRAM bit-cell. This is because while the CMOS transistor has undergone unprecedented scaling over the past few decades,  the sizes for state-of-the-art optical devices like photodiodes and MRRs are typically in the range of micro-meters \cite{de2021high, de2021high}. Thus, there exists a crucial non-trivial trade-off for O-SRAMs compared to E-SRAMs, wherein O-SRAMs can provide orders of magnitude faster speed and high energy efficiency while occupying significantly larger areas compared to E-SRAMs.
In this work, we aim to quantify the system-level benefits of O-SRAMs keeping in view the performance-power and area trade-off. For this work, we assume that O-SRAM acts as the on-chip memory for a wafer-scale FPGA, while the processing engines are built using conventional CMOS technology. While proposals for a high-speed optical logic circuit can be found in the literature \cite{xu2007all, singh2014all}, the scalability, noise resilience, and programmability of CMOS logic circuits are currently beyond the capability of optical logic circuits. Thus, our wafer-scale system is a heterogeneous system consisting of silicon photonics-based optical memories and CMOS-based processing engines. Note, due to the use of well-established silicon photonics devices for O-SRAM, the optical memory system can be seamlessly fabricated on the same wafer consisting of silicon CMOS transistors. 

\section{Performance Modeling Optical SRAM Memory} \label{perf_opt_mem}

\subsection{Modeling the operation of O-SRAM}

Fig. \ref{o-sram} shows the high-level view of an O-SRAM block. An O-SRAM block consists of optical storage, an optical-to-electrical conversion unit, and an electrical-to-optical conversion unit. Note that the electrical-to-optical conversion unit is not explicitly shown in Fig. \ref{o-sram} since it can be easily implemented using an active micro-ring resonator \cite{dong2009low}. The O-SRAM, in this case, operates at 20 GHz. Also, it can support multiple optical wavelengths (typically 5) through the use of wavelength division multiplexing, which enables concurrent access to the same  O-SRAM block. 

Fig. \ref{sram_pipes} illustrates the overall FPGA architecture with O-SRAM integrated. In the proposed work, O-SRAMs reside along with a configurable electrical mesh architecture. An O-SRAM uses a synchronization interface to connect with the configurable mesh due to the operation frequency difference between electrical compute components (i.e., electrical LUTs and DSPs) and optical memory components (O-SRAMs). 

In our work, we consider a FPGA with electrical memory components (i.e., electrical Block RAMs \cite{bram_xilinx} and Ultra RAMs \cite{uram_xilinx}) replaced by the same amount of O-SRAM memory. As illustrated in Figure \ref{sram_pipes}, A single O-SRAM can store 32 Kb of data. It contains 1024 data lines, where each data line has a width of 32 bits. Also, each O-SRAM consists of 200 parallel read-write ports with 32 bit-width as O-SRAMs support multiple optical wavelengths with high operating frequency.

For an O-SRAM that runs at $f_{\text{optical}}$ frequency using $\lambda$ number of wavelengths where each read-write port has a width of $z$ bits, the number of bits ($b_{\text{process}}$) it can provide to the electrical compute elements which runs at $f_{\text{electrical}}$ frequency is:

\vspace{-5mm}
\begin{equation}
  b_{\text{process}} = \dfrac{\lambda \times f_{\text{optical}} \times z}{f_{\text{electrical}}}
\end{equation}

Also, we focus on large tensor datasets where inputs initially reside in the FPGA external memory. FPGA external memory contains multiple DRAMs which use DDR4 technology.

\subsection{Modeling the energy consumption of O-SRAM}

The total energy consumption of an FPGA accelerator design ($E_{\text{FPGA}}$) on an O-SRAM-based FPGA is calculated by:

\vspace{-6mm}
\begin{multline}
    E_{\text{FPGA}} = P_{\text{compute}}\times t_{\text{runtime}} + E_{\text{DRAM-FPGA}} \\ + (P_{\text{O-SRAM}} \times n_{\text{O-SRAM}}) \times t_{\text{runtime}}
\end{multline}

Here, $P_{compute}$, $E_{\text{DRAM-FPGA}}$,  $n_{\text{O-SRAM}}$, $P_{\text{O-SRAM}}$, and $t_{\text{runtime}}$ refer to power consumption of compute resource of the FPGA, total energy consumption of DRAM-FPGA interface during external memory transactions, number of O-SRAMs used by the accelerator design, power consumption of a single O-SRAM, and run time of the accelerator, respectively.

The power consumption of a single O-SRAM block ($P_{\text{O-SRAM}}$) depends on the static power ($P_{\text{O-SRAM}}^{\text{static}}$ ) and switching power consumption ($P_{\text{O-SRAM}}^{\text{switching}}$) of the memory block. Static Power is primarily depends on the size of the SRAM ($S_{\text{O-SRAM}}^{\text{total}}$), electrical power per bit ($\hat{p}_{\text{optical storage}}$) and optical static power consumption per bit ($\hat{p}_{\text{optical}}^{\text{static}}$). $\hat{p}_{\text{optical storage}}$ and $\hat{p}_{\text{optical}}^{\text{static}}$ are a result of leakage power of optical and electrical components inside an O-SRAM. The switching power of O-SRAM ($P_{\text{O-SRAM}}^{\text{switching}}$) depicts the power consumed by O-SRAM during a read or write operation. It depends on the active number of bits of the O-SRAM in a given clock cycle ($S_{\text{O-SRAM}}^{\text{active}}$) following the power consumption of optical-electrical conversion per bit ($\hat{p}_{\text{optical-electrical conversion}}$) and per bit power consumption of optical storage units ($\hat{p}_{\text{optical storage}}$) shown in Fig. \ref{o-sram}(a). Equation \ref{eqn:power123} summarizes the O-SRAM power calculations.

\vspace{-4mm}
\begin{equation}
    \begin{split}
        P_{\text{O-SRAM}} & =   P_{\text{O-SRAM}}^{\text{static}} +P_{\text{O-SRAM}}^{\text{switching}} \\
        P_{\text{O-SRAM}}^{\text{static}} & = S_{\text{O-SRAM}}^{\text{total}}\times (\hat{p}_{\text{optical}}^{\text{static}} + \hat{p}_{\text{electrical}}^{\text{static}}) \\
        P_{\text{O-SRAM}}^{\text{switching}}  & = S_{\text{O-SRAM}}^{\text{active}}\times (\hat{p}_{\text{optical-electrical conversion}} + \hat{p}_{\text{optical storage}})
    \end{split}
    \label{eqn:power123}
\end{equation}



\section{\MakeLowercase{sp}MTTKRP Accelerator}

\subsection{spMTTKRP Computation} \label{mttkrp_access_patterns}

\begin{algorithm}
\DontPrintSemicolon
Input: A sparse tensor $\mathcal{X} \in \mathbb{R}^{I_0 \times I_1 \times I_2}$, dense factor matrices $\mathbf{{B}} \in \mathbb{R}^{I_1 \times R}$, $\mathbf{{C}} \in \mathbb{R}^{I_2 \times R}$ \;
Output: Updated dense factor matrix $\mathbf{A} \in \mathbb{R}^{I_0 \times R}$ \;

\For{each $i_0$ output factor matrix row in $\mathbf{A}$}{
$\mathbf{A}(i_0, :) = 0 $ \;
\For{each nonzero element in $\mathcal{X}$ at $(i_0,i_1,i_2)$ with $i_0$ coordinates}{
    Load($\mathcal{X}(i_0, i_1, i_2)$) \;
    Load($\mathbf{B}(i_1,:)$) \;
    Load($\mathbf{C}(i_2,:)$) \;
    \For{$r=1, \ldots, R$}{
        $\mathbf{A}(i_0, r) += \mathcal{X}(i_0, i_1, i_2) \times \mathbf{B}(i_1,r) \times \mathbf{C}(i_2,r)$ \;
    }
}
Store($\mathbf{A}(i_0,:)$) \;
}
\Return $\mathbf{{A}}$
\caption{{\sc spMTTKRP operation for mode 0 of a tensor with 3 modes}}
\label{mttkrp_appr_1_1}
\end{algorithm}

We use a hypergraph model to explain the spMTTKRP operation on a given input tensor. For illustrative purposes, we consider a 3 mode sparse tensor $\mathcal{X} \in \mathbb{R}^{I_0 \times I_1  \times I_2}$ where $(i_0, i_1, i_2)$ denote the coordinates of tensor element $x$ in $\mathcal{X}$. Here, $I_0$, $I_1$, and $I_2$ represent the size of each tensor mode. Note that the following approach can be applied to tensors with any number of modes.

\begin{figure}
\centering
\includegraphics[width=\linewidth]{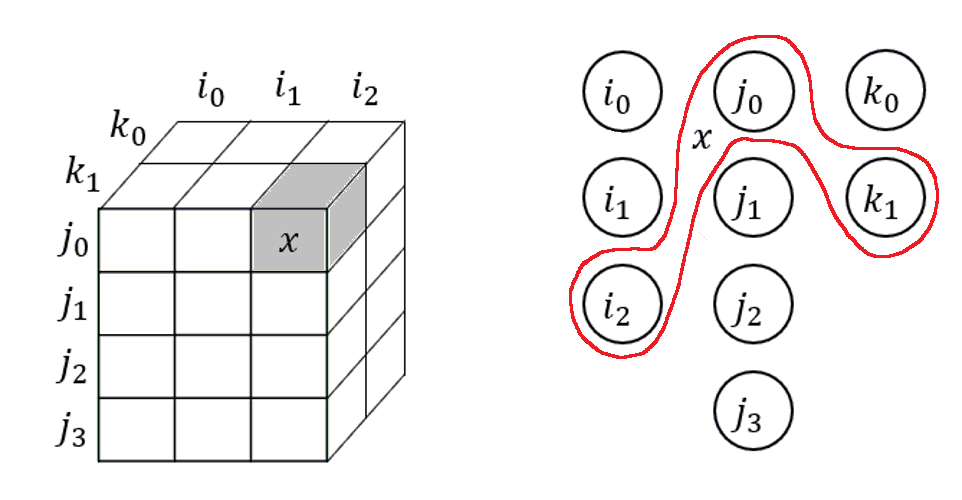}
\caption{A hypergraph example of a sparse tensor}
\label{hypergraph}
\vspace{-6mm}
\end{figure}

For a given tensor $\mathcal{X}$, we can build a hypergraph $H = (V, E)$ with the vertex set $V$ and the hyperedge set $E$ as follows: vertices correspond to the tensor indices in all the modes and hyperedges represent its non-zero elements. For a 3D sparse tensor $\mathcal{X} \in \mathbb{R}^{I_0 \times I_1 \times I_2}$ with $M$ non-zero elements, its hypergraph $H = (V,E)$ consists of $|V| = I_0 + I_1 + I_2$ vertices and $|E| = M$ hyperedges. A hyperedge $\mathcal{X}(i, j, k)$ connects the three vertices $i$, $j$, and $k$, which correspond to the indices of rows of the factor matrices. Fig.~\ref{hypergraph} shows an example of the hypergraph for a sparse tensor. 

Our goal is to determine a mapping of $\mathcal{X}$ into memory for each mode so that the total time spent on (1) loading tensor data from external memory, (2) loading input factor matrix data from the external memory, (3) storing output factor matrix data to the external memory, and (4) element-wise computation for each non-zero element of the tensor is minimized.

In Algorithm \ref{mttkrp_appr_1_1}, all hyperedges that share the same vertex of the output mode are accessed consecutively. The input vertices of the hyperedge are traversed to access rows of the remaining input factor matrices. It follows the element-wise multiplication and addition. Since the order of hyperedge depends on the output mode, the output factor matrix can be calculated without generating intermediate partial sums (Algorithm~\ref{mttkrp_appr_1_1}: line 10).
\\
\textbf{The total computations of the approach:} For a general sparse tensor with $|T|$ non-zero elements, $N$ modes, and factor matrices with rank $R$, since every hyperedge will be traversed once, and there are $N-1$ multiplication and one addition for computing spMTTKRP, the total computation per mode is $N \times |T| \times R$.
\\
\textbf{The total external memory (i.e., DRAMs) accesses:} It requires $|T|$ load operations for all the hyperedges and the total factor matrix elements transferred per mode is $(N-1) \times |T| \times R$, which corresponds to accessing input factor matrices of vertices in the hypergraph model. Let $I_{out}$ represent the length of the output mode. Then the total amount of data transferred is $|T| + (N-1) \times |T| \times R + I_{out}\times R$.

\begin{figure}
\centering
\includegraphics[width=0.7\linewidth]{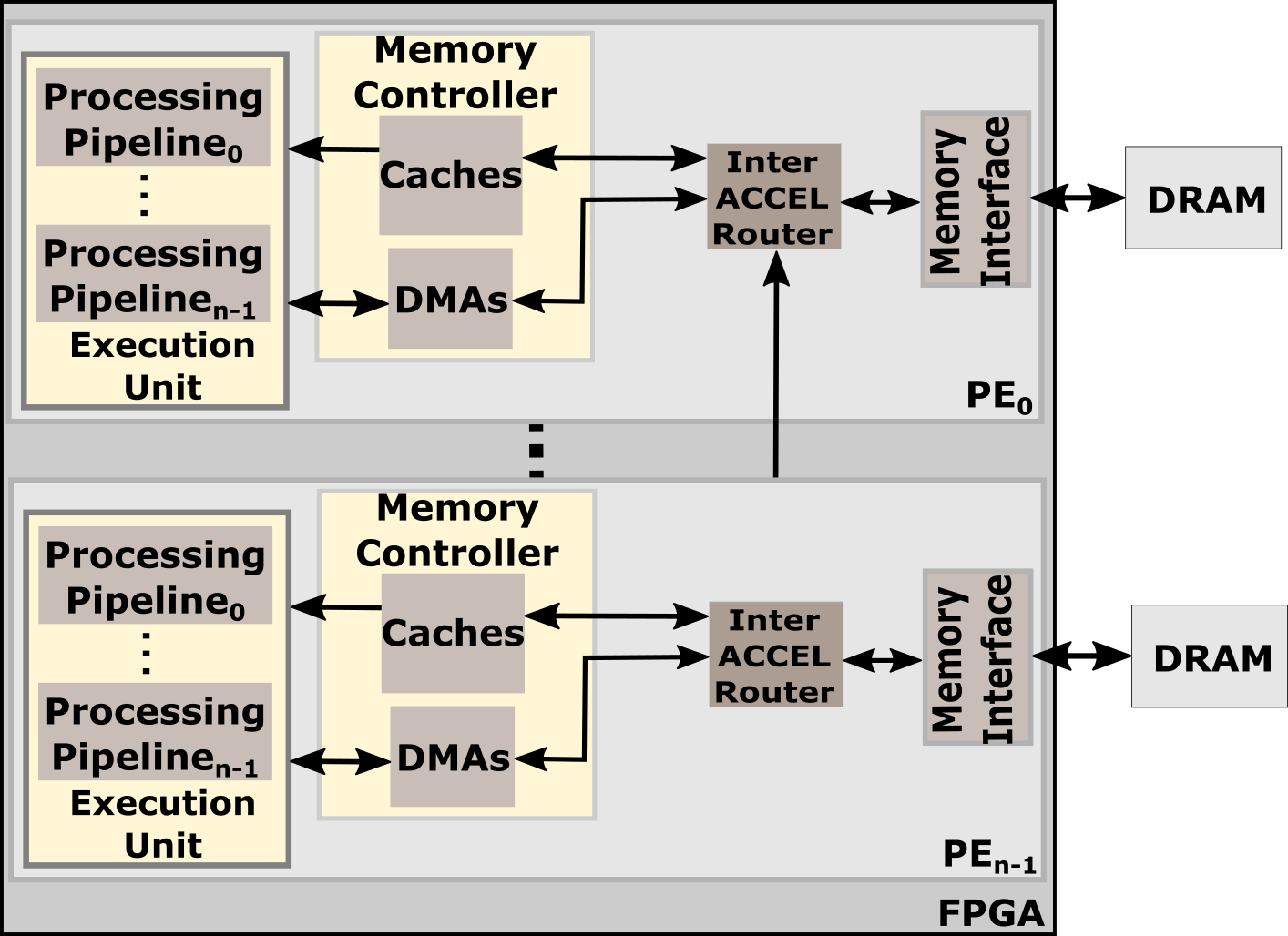}
\caption{Overall Architecture}
\label{overall}
\vspace{-4mm}
\end{figure}

The proposed sparse spMTTKRP computation has 4 main actions: 
(1) load a non-zero tensor element, 
(2) load corresponding factor matrices, 
(3) perform spMTTKRP operation, and 
(4) store the final output. 

We use a memory controller to decrease the total DRAM memory access time. It supports following access types:

\begin{enumerate}
  \item \textbf{Cache transfers}: Memory controller contains O-SRAM based multiple caches that Support random memory accesses. Load/store individual requests in minimum latency. Access patterns with high spatial and temporal locality are transferred using cache lines.
  \item \textbf{DMA stream transfers}: Memory controller contain SRAM based multiple Direct Memory Access (DMAs) that Support streaming accesses. Load/store operations on all requested data with minimum latency from memory.
  \item \textbf{DMA element-wise transfers}: DMAs can also be used to access data with no spatial and temporal locality.
\end{enumerate}

\subsection{Proposed FPGA Accelerator Design}


Fig. \ref{overall} shows the overall architecture of our FPGA accelerator design. We keep the number of Processing Elements (PEs) equal to the number of DRAMs attached to the FPGA. PE consists of a memory controller, execution unit, and external memory interface.
The execution unit inside the PE consists of parallel pipelines. It is a simple pipeline structure where computation demonstrated in Algorithm \ref{mttkrp_appr_1_1} is executed. Here, all the partial sums of spMTTKRP are stored inside an O-SRAM-based partial sum buffer.

\begin{figure}
\centering
\includegraphics[width=0.75\linewidth]{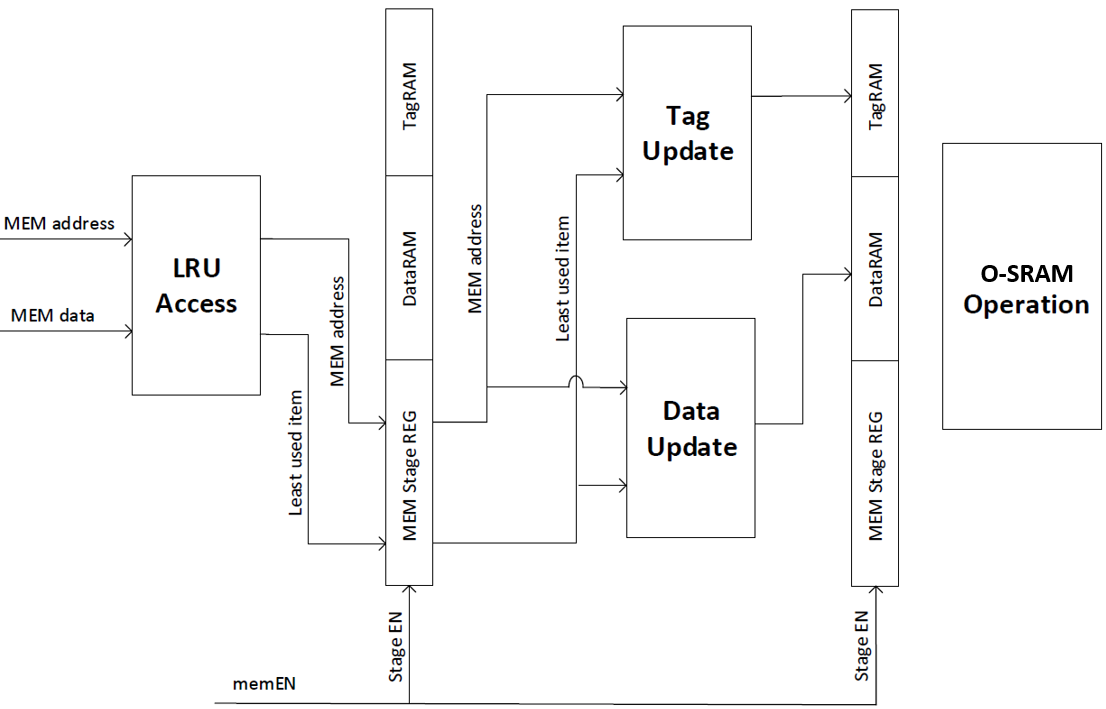}
\caption{Memory (MEM) pipeline of the cache}
\label{mem_pipe}
\vspace{-5mm}
\end{figure}

\begin{figure}
\centering
\includegraphics[width=0.75\linewidth]{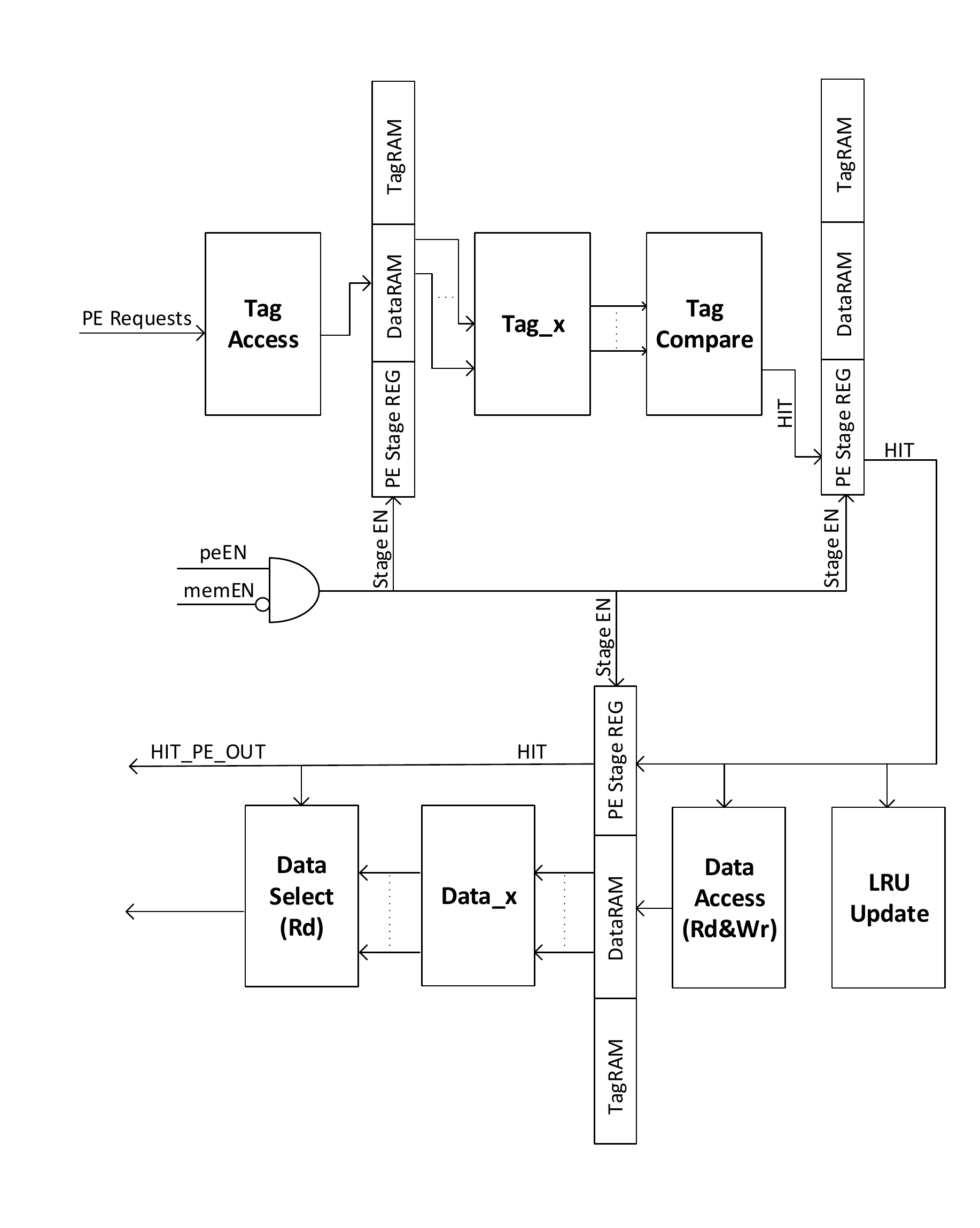}
\caption{Process element (PE) pipeline of the cache}
\label{pe_pipe}
\vspace{-5mm}
\end{figure}

The cache subsystem includes multiple O-SRAM-based caches. Each cache is shared with multiple input factor matrices. Each cache focuses on satisfying a single memory request with minimum latency. The cache uses two separate pipelines namely the PE pipeline and memory pipeline to support the high data rate of O-SRAMs due to high frequency. Fig. \ref{mem_pipe} and Fig. \ref{pe_pipe} depict the memory pipeline (MEM pipeline) and PE pipeline, respectively. They share the same Tag RAM, Data RAM, and LRU which are implemented using O-SRAMs. The PE pipeline is made into four stages, starting with a tag access step. Based on the address of the PE requests, tags are pulled out from the Tag RAM, denoted as \textit{Tag$\_$x}, and then compared to the incoming tag in the next stage. After the Tag comparator, cache hit information is generated and sent into the third stage. In this stage, the HIT information will be used as an evaluation criterion on whether the LRU update is needed or not. For read requests of \textit{m} (associativity) number of data, notated as \textit{Data$\_$x}, the data is pulled out from the Data RAM at the same time. Otherwise, for a written request with a hit, the updated data will be written into the corresponding entry of the Data RAM.

Each PE also contains Direct Memory Access units (DMAs) to access large tensors from the external memory. These DMAs have large buffers implemented using O-SRAMs.

\section{Evaluation}



\subsection{Experiments Setup} \label{ex_setup}

As described in section \ref{perf_opt_mem}, we consider a wafer-scale FPGA platform developed using 12 nm technology with O-SRAM as its internal memory. It contains a total of 54 MB of O-SRAM memory replacing the electrical SRAMs (E-SRAM) in a typical data center FPGA (eg., Xilinx Alveo U250) \cite{xilinxalveo}. It also contains 6433K LUTs, and 8474K Flip-flops, with 31K DSPs.
The compute units (configurable mesh) operate at 500 MHz, simulating a typical FPGA setup. Meanwhile, the O-SRAMs operate at 20 GHz as described in Section \ref{O-technology}.

The power and performance estimates for O-SRAM were obtained from electro-optic simulations using \textit{Lumerical Interconnect} \cite{Lumerical-pcie}. Estimates for E-SRAM were based on SRAM design in Globalfoundries 12nm node, compute and PE array primitives were synthesized to obtain power-performance-area estimates at 12nm Globalfoundries PDK. Finally, SPICE simulations were used to obtain the energy estimate for the optical-to-electrical interface.

\begin{table}[!ht]
\caption{Configurations of the accelerator}
\begin{center}
\resizebox{\columnwidth}{!}{%
\begingroup
\setlength{\tabcolsep}{6pt} 
\renewcommand{\arraystretch}{1.5} 
\begin{tabular}{|c|l|}
\hline
\textbf{Module} & \textbf{Configuration} \\
\hline
PE & Number of PEs: 4 \\
\hline
\multirow{2}{*}{
\begin{tabular}{c}
Parallel Pipelines
\end{tabular}} & No. of pipelines: 80 \\
\cline{2-2}
 & Partial Matrix Buffer size: 1024 elements \\
\hline
\multirow{5}{*}{
\begin{tabular}{c}
Cache sub system
\end{tabular}} & Number of caches: 3 \\
\cline{2-2}
& Associativity: 4 \\
\cline{2-2}
& Number of cachelines: 4096 \\
\cline{2-2}
& cachelines width: 64 B \\
\hline
\multirow{2}{*}{
\begin{tabular}{c}
DMAs
\end{tabular}} & No. DMA buffers: 6 \\
\cline{2-2}
& DMA buffer size: 64 KB  \\
\hline
\end{tabular}
\endgroup
}
\label{table1}
\end{center}
\vspace{-5mm}
\end{table}

\subsubsection{Dataset} \label{dataset}
We use the sparse tensors, derived from real-world applications, that appear in Table \ref{table3}. All of these tensors are provided by The Formidable Repository of Open Sparse Tensors and Tools (FROSTT) dataset~\cite{frosttdataset}. The selected dataset exhibits a variety of different sparse tensors in terms of dimensions, size, and density.

\begin{table}
\caption{Characteristics of the targeted sparse tensors}
\begin{center}
\resizebox{\columnwidth}{!}{%
\begingroup
\setlength{\tabcolsep}{6pt} 
\renewcommand{\arraystretch}{1.5} 
\begin{tabular}{ |c|c|c|c|c| }
 \hline
 \textbf{Tensor} & \textbf{Dimensions} & \#\textbf{NNZs} & \textbf{Density} \\
 \hline\hline
 NELL-1 & $2.9M \times 2.1M \times 25.5M$ & $143.6M$ & $9.1 \times 10^{-13}$ \\ 
 \hline
 NELL-2 & $12.1 K \times 9.2K \times 28.8K$ & $76.9M$ & $2.4 \times 10^{-05}$ \\ 
 \hline
 PATENTS & $46 \times 239.2K \times 239.2K$ & $3.6B$  & $1.4 \times 10^{-03}$ \\
 \hline
 LBNL & $1.6K \times 4.2K \times 1.6K \times 4.2K \times$ $868.1K$ & $1.7M$  & $4.2 \times 10^{-14}$ \\
 \hline
 DELICIOUS & $532.9K \times 17.3M \times 2.5M \times 1.4K$ & $140.1M$  & $4.3 \times 10^{-15}$ \\
 \hline
  AMAZON & $4.8M \times 1.8M \times 1.8M$  & $1.7B$ & $1.1 \times 10^{-10}$ \\
 \hline
  REDDIT & $8.2M \times 177K \times 8.1M$  & $4.7B$ & $4.0 \times 10^{-10}$ \\
 \hline
\end{tabular}%
\endgroup
}
\label{table3}
\end{center}
\vspace{-5mm}
\end{table}

\subsubsection{Implementation}

Table \ref{table1} shows the configuration parameters and the configuration we used in our experiments. Also, the tensor rank (R) is set to $16$ \cite{li2018hicoo}.

\subsubsection{Baseline platform} \label{baseline}
We consider the same FPGA with E-SRAM as the internal memory (i.e., electrical Ultra RAM \cite{uram_xilinx} and Block RAM \cite{bram_xilinx}) for the baseline.

\subsection{Overall Execution Time Performance}

\begin{figure}
\centering
\includegraphics[width=\linewidth]{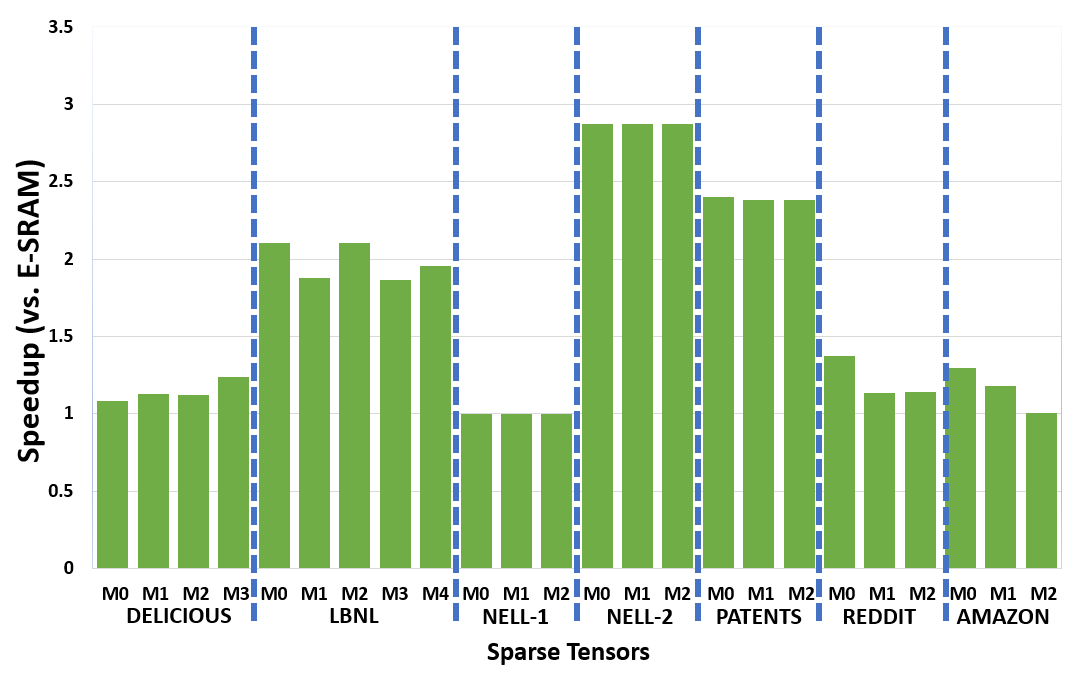}
\caption{Speedup achieved by replacing E-SRAM with O-SRAMs}
\label{speedup}
\vspace{-5mm}
\end{figure}

Fig.~\ref{speedup} shows the speedup the accelerator design achieved while using O-SRAMs. The baseline environment used for comparison is described in Section \ref{baseline}. In the horizontal axis, M$i$ ($i \in \{0,1,2,3,4\})$ refer to the each mode of the input tensor.

According to experiments, due to high data locality in memory accesses, NELL-2 and PATENT shows significant speedup while using O-SRAM. The support for concurrent accesses and the high clock frequency of O-SRAM contribute to this substantial speedup. Also, NELL-1 and DELICIOUS do not show significant speed up with O-SRAM as external FPGA memory access dominates the total execution time in these datasets.

\begin{figure}
\centering
\includegraphics[width=\linewidth]{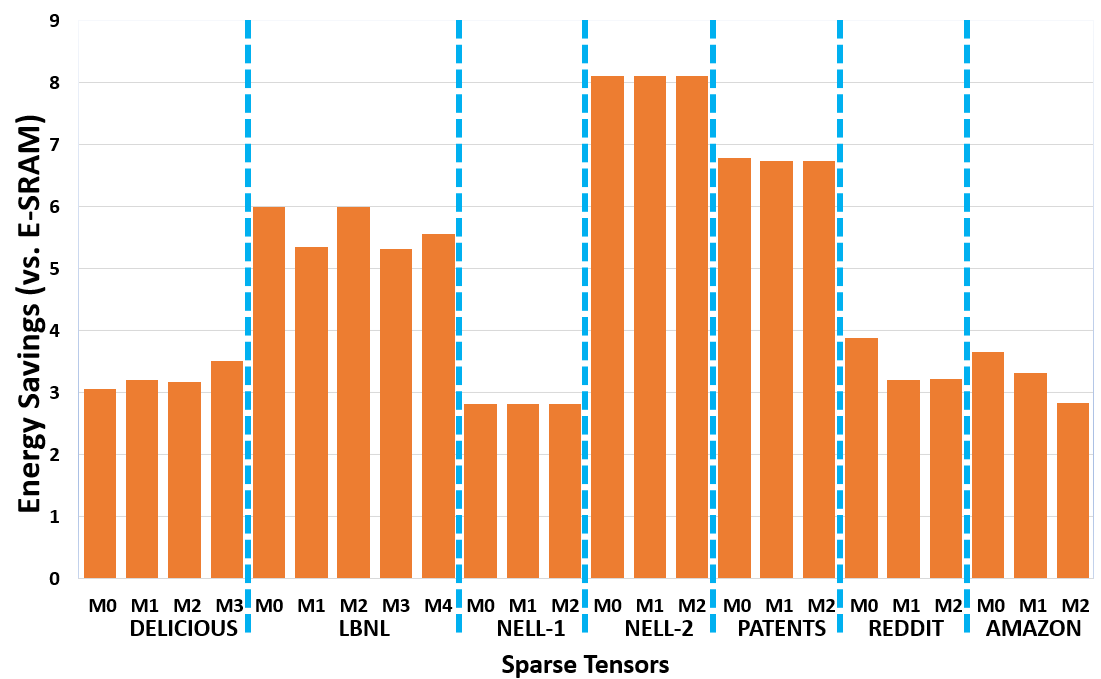}
\caption{Energy Savings by using O-SRAM technology}
\label{fig_energy}
\vspace{-1mm}
\end{figure}

\subsection{Overall Energy Performance}

\begin{table}
\centering
\caption{Energy consumption of the memory devices while FPGA operating at 500 MHz}
\begin{tabular}{c|c|c|c}
\hline
\multicolumn{4}{c}{\textbf{ Per bit Energy Consumption (pJ/cycle)}}                     \\
\hline
\multicolumn{2}{c|}{\textbf{Static}} & \multicolumn{2}{c}{\textbf{Switching}} \\
\hline
\textbf{Electrical}        & \textbf{Optical}        & \textbf{Electrical}          & \textbf{Optical}         \\
\textbf{Technology}        & \textbf{Technology}        & \textbf{Technology}          & \textbf{Technology}         \\
\hline
$1.175\times 10^{-6}$      &   $4.17\times 10^{-6}$        &       $4.68$      &     $1.04$     \\
\hline
\end{tabular}
\label{tab:Energy}
\end{table}

\begin{table}
\centering
\caption{The area with different SRAM technologies}
\begin{adjustbox}{max width=0.47\textwidth}
\begin{tabular}{|c|c|c|c|}
\hline
              & On-chip Memory & PEs & Total \\
\hline
E-SRAM system &   \begin{tabular}[|c|]{@{}c@{}} $43.2 ~mm^2$ \end{tabular}             &   $202.2 ~mm^2$       &   $247.2 ~mm^2$     \\
\hline
O-SRAM system & \begin{tabular}[|c|]{@{}c@{}} $103.7\times 10^4 ~mm^2$  \end{tabular}            &   $202.2 ~mm^2$         &   $103.7 \times 10^4 ~mm^2$   \\ 
\hline
\end{tabular}
\end{adjustbox}
\label{tab:area}
\vspace{-5mm}
\end{table}

The energy consumption of two memory technologies is shown in Table \ref{tab:Energy}. Compared with the electrical memory device, the optical memory device has lower switching energy driven by ultra-high operating frequency.

To compare the energy efficiency, the two FPGAs execute the same accelerator design with the datasets mentioned in Section \ref{dataset}. The energy comparison of  two FPGAs (i.e., O-SRAM-based FPGA and E-SRAM based FPGA) on different datasets is shown in Fig. \ref{fig_energy}.  According to the experiments results the O-SRAM FPGA is 2.8$\times$ - 8.1$\times$ more energy efficient than the E-SRAM FPGA.

\subsection{Area Comparison}
Although optical memory has shown low energy consumption with faster execution time, it occupies a significant area compared with the electrical, limiting the density of the optical memory and making large area wafer-scale systems a necessity for practical use of optical memory. Table \ref{tab:area} shows a breakdown of the area comparison between O-SRAM-based FPGA and E-SRAM-based FPGA.







\section{Conclusion}
In this paper, we perform comprehensive performance and energy consumption modeling of an optical SRAM-based FPGA environment for a sparse MTTKRP accelerator. The evaluation results show that using state-of-the-art optical memory technology, a FPGA can achieve an average of 1.68$\times$ speedup in execution time while saving an average of 5.3$\times$ more energy compared to an electrical SRAM-based system. Our future work includes reducing the area consumption of optical SRAM through multi-bit storage and optical device optimization.

\bibliographystyle{IEEEtran}
\bibliography{hpec20}

\end{document}